\documentclass[twocolumn,showpacs,preprintnumbers,amsmath,amssymb,prb]{revtex4}
\usepackage{graphicx,amssymb,amsmath}
\usepackage[dvips,usenames]{color}
\newcommand{\etalter}{{\it et al.}}
\newcommand{\eg}{{\it e.\ g.}}
\newcommand{\ie}{{\it i.\ e.}}
\begin{document}

\title{Calculation of Dynamical and Many-Body Observables with the Polynomial Expansion Method for Spin-Fermion Models}
\author{G. Alvarez}
\author{T. C. Schulthess}
\affiliation{Computer Science \& Mathematics 
Division, Oak Ridge National Laboratory, \mbox{Oak Ridge, TN 37831}}

\begin{abstract}

The calculation of two- and four-particle observables is addressed within the framework of the truncated 
polynomial expansion method (TPEM). 
The TPEM
replaces the exact diagonalization of the one-electron sector in models for fermions
coupled to classical fields such as those used in manganites and diluted magnetic semiconductors. The computational
cost of the algorithm is $O(N)$ -- with $N$ the number of lattice sites --
 for the TPEM which should be 
contrasted with the
computational cost of the diagonalization technique that scales as
$O(N^4)$. By means of the TPEM, the density of states, spectral function and optical conductivity of a double-exchange
model for manganites are calculated on 
large lattices and 
compared to previous results and experimental measurements. 
The ferromagnetic metal becomes an insulator by increasing the direct exchange coupling that competes with
the double exchange mechanism. This metal-insulator transition is investigated in three dimensions. 
\end{abstract}
 
\maketitle

\section{Introduction}

Strongly correlated electron systems continue to be one the most important areas of research in condensed matter physics. 
In this context,
spin-fermion models, \ie\ fermion systems coupled to classical fields, provide a bridge between the full Hubbard model 
and models with only
classical degrees of freedom. The later fail to capture important physical aspects, like carrier mediated ferromagnetism in manganites
while the former are usually too complex to study without approximations due to the sign problem.\cite{re:schmidt84} 

Spin-fermion models contain
 all the physical properties of a 
strongly correlated electron system, including the effect of the carriers and inhomogeneous local charge density.  They 
have been successful in the description of
many materials, \eg\ manganites\cite{re:dagotto02} and diluted magnetic semiconductors (DMS).\cite{re:alvarez03} 
Inhomogeneous phases in these materials have been found to be of importance in explaining many physical properties.
It was also found that mean field theories are insufficient to obtain these phases in DMS 
systems\cite{re:kaminski02,re:alvarez02} but that information from these models
must be extracted without approximations. In 2005, a spin-fermion model for multiferroic materials was also 
studied with the diagonalization method.\cite{re:sergienko05} 

 The traditional method to calculate observables has been to integrate the fermions exactly
at a finite temperature for every configuration of classical fields and then integrate these fields by means of a Monte Carlo
procedure. Although this method -- which will be referred to as DIAG -- is accurate and has lead to many advances, \eg\ the theoretical observation of phase 
separation in manganites\cite{re:dagotto02} and
clustered states in DMS,\cite{re:alvarez03} it has limitations. In fact, the largest systems that can be studied with this
procedure are lattices with less than approximately 300 sites.

In 1999, Furukawa and coworkers\cite{re:motome99} introduced a polynomial expansion of the fermionic density of states as a way of avoiding
the diagonalization of the fermionic sector. This polynomial expansion method (PEM) scales as $O(N^3)$ as opposed to the diagonalization method that
scales as $O(N^4)$, $N$ being the size of the system. The method could also be parallelized, which is practically impossible for the
diagonalization method. In 2003, the same researchers\cite{re:furukawa03} developed a truncated version of the PEM, that will be abbreviated by TPEM that
scales linearly, \ie\ $O(N)$, with the size of the system. All these methods do not introduce systematic errors or 
uncontrolled approximations and
the results obtained converge to the exact result as the expansion cut-off tends to infinity.

Although the TPEM is becoming a standard way to avoid the 
diagonalization,\cite{re:aliaga05,re:motome03b,re:dobrovitski03,re:wang94} observables that involve two-
and four- fermionic operators have not been studied systematically. Here, it will be shown how to efficiently calculate these observables
and the formalism is applied to a double exchange model for manganites. The emphasis will be on the calculation of the spectral
function,
$A(k,\omega)$, and optical conductivity, $\sigma(\omega)$, although the treatment applies to other observables as well.

Another algorithm that improves the integration of the one-electron sector in double exchange models is the ``Hybrid Monte Carlo'' (HMC)
algorithm.\cite{re:alonso01} This method uses the path integral representation, introducing imaginary time and a HMC
procedure to evolve the system. The TPEM works better than the HMC at low temperatures where the HMC presents increasing computational cost due
to the time discretization. Another reason why we have preferred the TPEM is that the HMC method
is applicable when the bands are connected and do not extend over a wide range of energies,
as is the case of finite coupling systems. The TPEM also allows for easy parallelization, improving the performance even more.

The contents of this paper is the following.
In \S\ref{subsec:review}  a review of the TPEM for spin fermion models as introduced originally by Furukawa \etalter\ is presented. In
\S\ref{subsec:dos}-\ref{subsec:twoparticle} the formalism to calculate two-particle observables is discussed as well as examples and in particular the
formula for the spectral function, $A(k,\omega)$. In \S\ref{subsec:fourparticle} the formalism is extended to include
four-particle observables and to allow for the calculation of the optical conductivity. 
The parallelization of the algorithm and its complexity is discussed in \S\ref{subsec:parallel}.
Then, in Section \ref{sec:results}
 these observables are calculated for a
model of manganites. The metal to insulator transition is studied in three dimensions.
 Finally, in Section \ref{sec:conclusions}, we present
the most important conclusions and discuss further uses of these techniques.
 
\section{Theory}

\subsection{Review of the TPEM}\label{subsec:review}
The brief review presented here follows closely Reference~\onlinecite{re:alvarez05}.
Consider a model defined by a general Hamiltonian,
$\hat{H}=\sum_{ij\alpha\beta}c^\dagger_{i\alpha} H_{i\alpha,j\beta}(\phi)c_{j\beta}$, 
where the indices $i$ and $j$ 
denote a spatial index and $\alpha$ and $\beta$ are internal degree(s) of freedom, \eg\ 
spin and/or orbital.
The Hamiltonian matrix depends on the configuration of one or more classical fields,
represented 
by $\phi$. Although no explicit indices will be used, the field(s) $\phi$ are supposed to have a spatial dependence. 
The partition function for this Hamiltonian is given by:
$
Z=\int d\phi \sum_n \langle n | \exp(-\beta\hat{H}(\phi)+\beta\mu\hat{N}) |n\rangle
$
where $|n\rangle$ are the eigenvectors of the one-electron sector. Here $\beta=1/(k_B T)$ is the inverse temperature.
The number of particles  (operator $\hat{N}$) is adjusted
via the chemical potential $\mu$. The procedure to calculate observables (energy, density, action, etc) is the following. 
First, an arbitrary configuration of classical fields $\phi$ is selected as a starting point.
The Boltzmann weight or action of that configuration, $S(\phi)$, is calculated by diagonalizing the one-electron sector at
finite temperature. Then a small local change of the field configuration is proposed, so that the new configuration
is denoted by $\phi'$ and its action is evaluated to obtain the difference in action $\Delta S = S(\phi')-S(\phi)$. 
Finally, this new configuration is accepted or rejected based on a
Monte Carlo algorithm like Metropolis or heat bath and the cycle starts again. 
In summary, the observables are traditionally 
obtained 
using exact 
diagonalization of the one-electron sector for every classical field configuration and Monte Carlo integration for those
classical fields \cite{re:dagotto02}. The PEM/TPEM replaces the diagonalization for a
polynomial expansion and is briefly described here but further details can be found in
Ref.~\onlinecite{re:motome99,re:furukawa01,re:furukawa03}. These definitions will be helpful later in 
deriving formulas for many-body observables.

It will be assumed that the Hamiltonian $H(\phi)$ is normalized, which simply implies a re-scaling:
\begin{eqnarray}
H&\rightarrow& (H-b)/a\nonumber\\
a&=&(E_{max}-E_{min})/2\nonumber\\
b&=&(E_{max}+E_{min})/2,\label{eq:hamnormalization}
\end{eqnarray}
where $E_{max}$ and $E_{min}$ are the maximum and minimum eigenvalues of the original Hamiltonian respectively.
This in turn implies that the normalized Hamiltonian has eigenvalues $\epsilon_v\in[-1,1]$.

Observables can be divided in three categories:
(i) those that do not depend directly on fermionic operators, \eg\ 
 the magnetization, susceptibility and classical spin-spin correlations 
 in the double
exchange model,
(ii) those for which a function $F(x)$ exists
such that they can be written as:
\begin{equation}
A(\phi)=\int_{-1}^{1}F(x)D(\phi,x)dx,
\label{eq:observable}
\end{equation}
where $D(\phi,\epsilon)=\sum_\nu\delta(\epsilon(\phi)-\epsilon_\nu)$, 
and $\epsilon_\nu$ are the eigenvalues of $H(\phi)$, \ie\ $D(\phi,x)$ is the density of
states of the system. Finally, category (iii) would include 
fermionic observables that cannot be expressed as in Eq.~(\ref{eq:observable}), in particular two and four-particle
observables. In 
this paper the emphasis will be on this last category, that will be analyzed in the remaining sections, 
since dynamical observables are expressed in terms of two-particle observables, 
\eg\ the spectral function and four-particle linear responses,
 \eg\  optical conductivity, that had not been studied before with the TPEM or on large systems. 
 
For category (i), the calculation is straightforward and simply involves an average over 
Monte Carlo configurations.
Category (ii) includes the effective action or generalized
Boltzmann weight and this quantity is particularly important because it is calculated very frequently throughout the 
Monte Carlo procedure to integrate the classical fields. Therefore, we first briefly review the calculation of these types of observables.
As Furukawa \etalter\ showed, it is convenient to expand a function $F(x)$ in terms of Chebyshev
polynomials in the following way:
\begin{equation}
F(x)=\sum_{m=0}^{\infty}f_mT_m(x),
\label{eq:coeffs}
\end{equation}
where $T_m(x)$ is the $m-$th Chebyshev polynomial evaluated at $x$.
Let $\alpha_m=2-\delta_{m,0}$. The coefficients $f_m$ are calculated with the formula:
\begin{equation}
f_m=\int_{-1}^1 \alpha_m F(x)T_m(x)/(\pi\sqrt{1-x^2}).
\label{eq:coeffscalculation}
\end{equation}
The moments of the density of states are defined by: 
$\mu_m(\phi)\equiv\sum_{\nu=1}^{N_{dim}}\langle \nu| T_m(H(\phi))|\nu\rangle,$
where $N_{dim}$ is the dimension of the one-electron sector.
Then, the observable corresponding to the function $F(x)$, can be calculated by:
$
A(\phi)=\sum_m f_m \mu_m(\phi).
$
In practice, the sum in this equation is performed up 
to a certain cutoff value
$M$, without appreciable loss in accuracy.\cite{re:motome99,re:furukawa01}
The calculation of $\mu_m$ is carried out recursively.
$|\nu;m\rangle  =  T_m(H)|\nu\rangle=2H|\nu;m-1\rangle-|\nu;m-2\rangle$
and hence 
$
\mu_{2m} =  \sum_\nu(\langle m;\nu|\nu;m\rangle -1 )\nonumber$ and
$\mu_{2m+1} = \sum_\nu(\langle m;\nu |\nu;m +1\rangle -\langle\nu;0|\nu;1\rangle),$
are used to calculate the moments.
The process involves a sparse matrix-vector product, \eg\ in $T_m(H)|\nu\rangle$,
 which, when considering the trace operation, yields a cost  of
$O(N^2)$ for each configuration, implying $O(N^3)$ for each Monte Carlo step. In addition,
an improvement of the present method has been proposed \cite{re:furukawa03} based on 
a truncation of the matrix-vector product mentioned before
and it turns out that the resulting algorithm has a complexity linear in $N$. This approximation
is controlled by the small parameter $\epsilon_{pr}$.

The function corresponding to the effective action for a configuration $\phi$ is defined by
$F^S(x) = -\log(1+\exp(-\beta(x-\mu)))$ and $S(\phi)$ admits an expansion as before with
coefficients $f^S_m$ corresponding to $F^S(x)$.
This observable is calculated very frequently in the Monte Carlo integration procedure and so it is important to
calculate it efficiently.
In actuality, only the difference in action,
$\Delta S = S(\phi')-S(\phi)$  has to be computed at
every change of classical fields. 
The authors of Ref.~\onlinecite{re:furukawa03} have also developed a truncation procedure for this
trace operation controlled by a small parameter, $\epsilon_{tr}$ and the current value of the
effective action can be calculated faster using the moments of the expansion from the
previous MC update. This truncation is based on the observation that if $\phi$ and $\phi'$ differ only in very few sites 
(as is the case with local Monte Carlo updates) then the
corresponding moments will differ only for certain indices. More details can be found in Ref.~\onlinecite{re:furukawa03}.

In what follows, the size of the Hilbert space will be denoted by $N_{dim}$ and it will depend on the size
 of the lattice as well as on the particular model to be studied. For a one-orbital
double exchange
model on a lattice of $N$ sites and finite coupling, $N_{dim}=2N$; the factor of 2 is due to the spin degree of freedom.

\subsection{Density-Of-States}\label{subsec:dos}
The fermionic density of states of the system for a configuration of classical fields $\phi$ is given by
$
N_\phi(\omega)= \sum_\lambda \delta(\omega'-\epsilon_\lambda),
$
where $\omega'=(\omega-b)/a$. Then, the corresponding function for $D_\phi(\omega)$ in Eq.~(\ref{eq:observable}) is
$F(x)=\delta(\omega'-x)$ and the corresponding fixed coefficients are calculated from Eq.~(\ref{eq:coeffscalculation}). The end result is
\begin{equation}
N_\phi(\omega)=\frac{\sum_m \alpha_m T_m(\omega') \mu_m(\phi)}{\pi \sqrt{1-\omega'^2}}.
\label{eq:tpemdos}
 \end{equation}
A Monte Carlo average over classical fields is generally needed to obtain the fermionic density of states, \ie, 
$N(\omega)=\langle N_\phi(\omega)\rangle_\phi$, and for that it suffices to calculate $\langle\mu_m(\phi)\rangle_\phi$, as can be
deduced from the form of the last equation.

The sum in Eq.~(\ref{eq:tpemdos}) is truncated to a certain cut-off $M$ (see, \eg\ Ref.~\onlinecite{re:furukawa01}) and
such an abrupt truncation generally results in unwanted Gibbs oscillations. 
Silver \etalter\  have considered\cite{re:silver94,re:silver96} instead smooth truncations, where the moments are multiplied
by damping factors, $g_m$. A possible choice of the damping factors, $g_m$, can be found in Ref.~\onlinecite{re:silver94,re:silver96}.

\subsection{Two-particle observables}\label{subsec:twoparticle}
The previous derivations can be slightly modified to allow for the calculation of two-particle observables. These can be written as:
\begin{equation}
A_{i\alpha,j\alpha'}(\phi)=\int_{-1}^{1} F(x) D_{i\alpha,j\alpha'}(\phi,x) dx,
\label{eq:dosexpansion}
\end{equation}
where 
\begin{equation}
 D_{i\alpha,j\alpha'}(\phi,x)=\sum_\lambda U^\dagger_{\lambda,i\alpha} U_{j\alpha',\lambda} \delta (x-\epsilon_\lambda)
\end{equation}
and $F(x)$ is a function that defines the observable in question. 
The quantities $ U_{j\alpha,\lambda}$ and $\epsilon_\lambda$ denote the eigenvectors and eigenvalues of 
the normalized one-electron Hamiltonian, $H(\phi)$,
respectively. (In general, $i$ and $j$ are spatial indices and $\alpha$ and $\alpha'$ are indices corresponding to internal degrees of
freedom if any. Spatial indices will always be denoted without arrows or bold letters independently of the dimension since on any
 lattice a vector ${\mathbf v}$ can be 
mapped into a single integer. Also the notation $i+j$ is meant to represent the lattice site given by the vectorial sum of the vectors corresponding to 
$i$ and $j$ respectively.)
For example, in the context of a one-orbital 
spin-fermion model with spin, consider the observable:
$
\langle c^\dagger_{i\sigma} c_{j\sigma}\rangle   = \sum_\lambda   U^\dagger_{\lambda,i\sigma} U_{j\sigma,\lambda} f(\beta
(a\epsilon_\lambda+b-\mu)),
$
with 
$
f(x)=\frac{1}{({\rm e}^x+1)}
$
 and the $F(x)$ of Eq.~(\ref{eq:dosexpansion}) is
  $F(x)=f(\beta (ax+b-\mu))$ for this particular observable.(The normalization factors $a$ and $b$ where defined
 in Eq.~(\ref{eq:hamnormalization}).) 
Now, $F(x)$ is expanded in Chebyshev
polynomials, $F(x)=\sum_{m=0}^{\infty} f_m T_m(x)$, and the coefficients $f_m$ are calculated by inverting this formula.
In all equations $\langle\cdots\rangle$ denotes the average over the fermionic sector for a fixed configuration of classical fields.
The moments of $D_{i\alpha,j\alpha'}$ are defined by:
\begin{equation}
\mu_{i\alpha,j\alpha';m}(\phi)=\langle i\alpha|T_m(H(\phi))|j\alpha'\rangle,
\label{eq:momij}
\end{equation}
and therefore:
\begin{equation}
A_{i\alpha,j\alpha'}(\phi)=\sum_m f_m \mu_{i\alpha,j\alpha';m}(\phi),
\end{equation}
which is the formula used for the calculation of the observable $A_{i\alpha,j\alpha'}$ with the TPEM. The computational complexity of the
calculation of $\mu_{i\alpha,j\alpha';m}$ can be inferred from the computational complexity of the calculation of the moments of the density
of states,\cite{re:furukawa03} and is $O(1)$ for TPEM and $O(N)$ for PEM. In general, non-local observables 
involve a sum over the index $i$ or
$j$ in the previous equation. For instance, the kinetic energy is written as $K= -t\sum_{\langle ij \rangle,\sigma} (c^\dagger_{i\sigma} c_{j\sigma} + {\rm H.c.})$
and, therefore, the complexities are multiplied by a factor of $N$. However, this still gives $O(N)$ complexity when the
truncated expansion is used. It is important to remind the reader that two-particle observables are not required at every
change of classical fields in 
 a Monte Carlo simulation but are computed at most only once per Monte Carlo step.
One observable of particular interest is the spectral function, $A(k,\omega)$ that can measured with
Angle Resolved Photoemission Spectroscopy (ARPES). 
This observable can be
 expressed in terms of the eigenvalues $\epsilon_\lambda$ and eigenvectors $U_{\lambda,i\sigma}$ of the normalized 
Hamiltonian\footnote{A fixed configuration of classical fields, $\phi$, is assumed. The quantities are later averaged over these classical fields
by the Monte Carlo procedure explained before. The explicit dependence on $\phi$ will be omitted whenever there is no  possibility of
confusion.}:
$
A_\phi(r,\omega)=\sum_{i,\sigma,\lambda} U^\dagger_{\lambda,i\sigma}(\phi) U_{(i+r)\sigma,\lambda}(\phi) \delta (\omega'-\epsilon_\lambda),
$ 
where $\omega'=(\omega-b)/a$ is the normalized frequency. Then, it is straightforward to show that the function $F(x)$ of
Eq.~(\ref{eq:dosexpansion}) for $A(r,\omega)$ is $F^A(x)=\delta (\omega'-x)$, which allows for the calculation of the
Chebyshev coefficients in terms of the moments of $D_{i\sigma,j\sigma}(\phi,x)$ with the help of Eq.~(\ref{eq:coeffscalculation}):
\begin{equation}
A(r,\omega)=\sum_m  \frac{\alpha_m T_m(\omega') \sum_{i\sigma}\mu_{i\sigma,(i+r)\sigma;m}}{\pi\sqrt{1-\omega'^2}}.
\label{eq:arw}
\end{equation}
(Note that $\alpha_m=2-\delta_{m,0}$ was defined previously.)

\subsection{Four-particle observables}\label{subsec:fourparticle}
Linear responses cannot in general be expressed in terms of only two-particle observables but they  can nevertheless be calculated with the
formalism of the TPEM without introducing random vectors, uncontrolled approximations or systematic errors, as will be shown.
 The discussion here will be focused on the optical
conductivity but the procedure can be generalized to other observables as long as there is an operator that plays the role of the current.
The key step is based on a recent development by
Weisse\cite{re:weisse04,re:weisse05} in the context of the Anderson model. 

For any configuration of classical fields $\phi$, the optical conductivity\cite{re:alvarez03}
can be rewritten as ($f(x)$ is the Fermi function):
$
\sigma(\omega)_\phi = \frac{\pi}{\omega N} \sum_{\lambda\lambda'} |\langle\lambda|j_x|\lambda'\rangle|^2   (f(\beta (E_{\lambda'}-\mu))- 
f(\beta (E_\lambda-\mu)) \delta(\omega - E_\lambda + E_{\lambda'}),
$
where $|\lambda\rangle$ and $E_\lambda$ denote the eigenvectors and eigenvalues of the one-electron Hamiltonian. Note that 
$E_\lambda$ is related to
the normalized Hamiltonian eigenvalues simply by $E_\lambda =
a\epsilon_\lambda +b$ (see Eq.~(\ref{eq:hamnormalization})).
Now, following Reference \onlinecite{re:weisse04}, $\sigma(\omega)$ can be expressed in terms of the function $j_\phi(x,y)$:
\begin{eqnarray}
j_\phi(x,y)&=&\frac1N \sum_{\lambda\lambda'}|\langle\lambda|j_x|\lambda'\rangle|^2 \delta(x-\epsilon_\lambda) \delta(y-\epsilon_{\lambda'})\\
\nonumber
\sigma_\phi(\omega)&=&\frac{1}{\omega}\int_{-\infty}^{\infty} j_\phi(x,x+\omega)\left[ f(\beta (ax+b-\mu))-\right. \\
& & \left. f(\beta (ax+b-\mu+\omega))\right] dx.
\end{eqnarray}
The function $j_\phi(x,y)$ can be expanded in Chebyshev polynomials:
$
j_\phi(x,y)=\sum_{mn} \eta_{mn}(\phi) T_m(x) T_n(x),
$
with
$
\eta_{mn}(\phi)\equiv{\rm Tr}(T_m(\hat{H}(\phi))j_xT_n(\hat{H}(\phi))j_x)/N
$.
If the above trace is evaluated in the real-space basis, $\{|i>,\sigma\}$, of the one-electron sector, then 
it is possible to express $\eta_{mn}$ in terms of the coefficients $\mu_{i\sigma,j\sigma';m}$ defined in Eq.~(\ref{eq:momij}):
\begin{eqnarray}\nonumber
\eta_{mn}(\phi)&=&\sum_{i,j,k,l}\sum_{\sigma_1,\sigma_2,\sigma_3,\sigma_4} \mu_{i\sigma_1,j\sigma_2;m}(\phi) \langle j\sigma_2|j_x|k\sigma_3\rangle \times \\
& & \times \mu_{k\sigma3,l\sigma4;n}(\phi) 
\langle l\sigma_4 |j_x|i\sigma_1\rangle,
\end{eqnarray}
and the elements  $\langle i\sigma|j_x|j\sigma'\rangle$ can be easily calculated: they are equal to $\pm ite \delta_{\sigma,\sigma'}$ if $j=i\pm \hat{x}$ or $0$ otherwise.
Now it is possible to calculate, $\eta_{mn}(\phi)$ for any classical field configuration $\phi$.
Finally,  a Monte Carlo averaging over the configurations $\phi$ is performed that yields $\sigma(\omega)$:
\begin{eqnarray}
\sigma(\omega)&=&\frac{1}{\omega}\int_{-\infty}^{\infty} 
\sum_{mn} \langle \eta_{mn}(\phi)\rangle_{\phi} T_m(x) T_n(x+\omega)  \times \nonumber\\
& & (f(\beta x)-f(\beta (x+\omega))) dx.
\end{eqnarray}
Note that the Monte Carlo procedure has to average only $M\times M$ quantities  $\langle \eta_{mn}(\phi)\rangle_{\phi}$ and the
computation of $\sigma(\omega)$ using the previous equation needs to be performed only once per Monte Carlo step.
It is also worth remarking that our method differs from the one proposed by Iitaka et al. \cite{re:iitaka97} because it
 does not introduce random vectors and it is
applicable at finite temperature.

\subsection{Parallel Computation} \label{subsec:parallel}
Another advantage of the TPEM, is that it can be parallelized effectively up to 50-100 processors,
 because 
 the calculation of the moments 
 for each basis
ket $|\nu\rangle$ is independent. Thus, the basis can be partitioned in such a way that each processor
calculates the moments corresponding to a portion of the basis. It is important to remark that the data to be moved between
different nodes are small compared to calculations in each node. Indeed, the communication time is
proportional to $MN_{PE}$ and communication among nodes is mainly done here to add up all the
moments.
The exact diagonalization algorithm can be parallelized only up to very few processors even when considering sparse matrices.

The calculation of observables that will be presented in the next section was carried out by using
parallel computation. The complexity scales with the number of processors up to the size of the
one-electron Hilbert space, $N_{dim}$. However, the observable that needs to be calculated more
frequently is the difference in effective action. Due to the truncation in this action difference that
takes place when the TPEM is used (see \eg\  
Ref.~\onlinecite{re:furukawa03}), the sum in this case is performed only up to the size of the truncated Hilbert
space, $A$. This implies that the inverse complexity scales with the
number of processors, $N_{PE}$ only up to $N_{PE}=A$ and not $N_{PE}=N_{dim}$. 
Although this imposes a limitation of the parallelization algorithm, since in general $A\ll
N_{dim}$, energy integrated observables usually converge for a cutoff $M\le20$ and less than 
$A$ processors is enough in most cases.\cite{re:alvarez05}.

However, energy dependent or many-body observables, \eg\ the density-of-states and the optical conductivity, need
more moments (usually 100 or more as will be discussed in the next section)
 for their calculation within the same error. Now, increasing the number of moments $M$
implies an increase of the truncated Hilbert space of size $A$. 
But even though now the
number of moments has to be increased, the threshold for the scaling of the parallelization algorithm has
also increased. Therefore, these types of problems are particularly suited 
to be studied efficiently with supercomputers with thousands of nodes.

\section{Application: Double Exchange Model For Manganites}\label{sec:results}
\subsection{Microscopic Model}

\begin{figure}[h]
\includegraphics[width=8cm,clip]{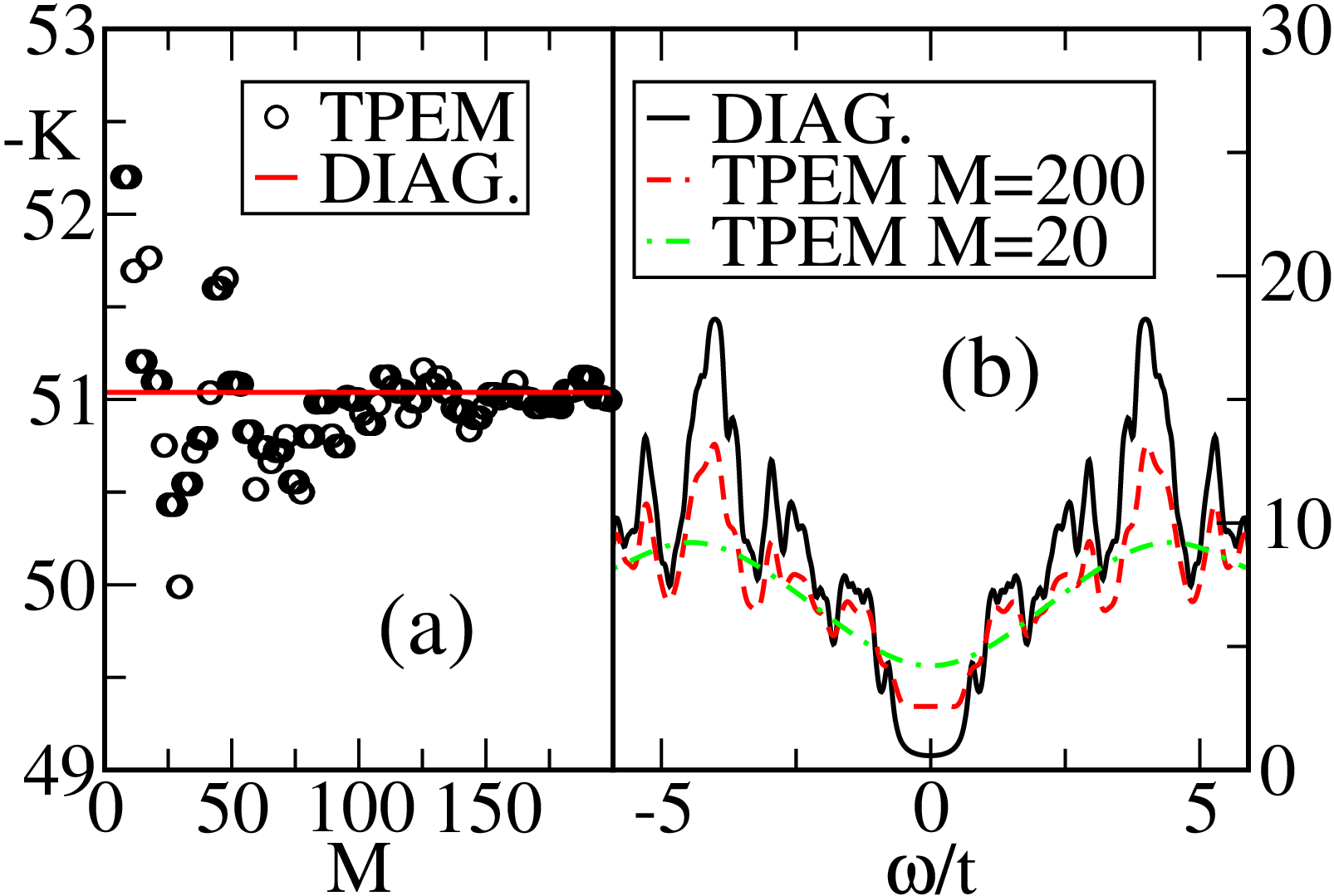}
\caption{(a) Kinetic energy of a single configuration of classical spins on a 8$\times$8 lattice for model Eq.~(\ref{eq:hammanganites}) 
calculated with
the DIAG and TPEM algorithms. The configuration was obtained after evolving the model with Monte Carlo with 600 steps for $J_H=4t$ and
$\mu=-4t$, $T=0.02t$ and $J_{AF}=0$. $\epsilon_{pr}=10^{-5}$ and $\epsilon_{tr}=10^{-6}$.
(b) Density-of-States for the model for the same configuration and parameters as in (a). 
Algorithms and different expansion cut-offs are indicated.
\label{fig:kinetic}}
\end{figure}

Manganese oxides, known as manganites, are  correlated-electron systems of much current interest in condensed matter physics. 
This interest started with the discovery of the so-called ``colossal"
mangetoresistance (CMR) effect, an effect three orders of magnitude larger
than the one observed previously on superlattice films.\cite{re:dagotto05}

Considerable progress has been achieved in understanding many properties
of these compounds by using double exchange models.\cite{re:dagotto01}
The electron-phonon coupling is usually neglected but it can and has been included; the phonon degrees of freedom are, then, an additional
classical field. In the ferromagnetic phase the electrons directly jump
from manganese to manganese spin and their kinetic energy is minimized when these spins are
aligned. Moreover, the unbiased study of spin-fermion models for manganites as presented here can produce other long- and short-range 
ordered phases,
\eg\ antiferromagnetism, inconmensaturate phases (to be described later) and phase separation. In general,
the correct ground states and very good estimates of Curie and Ne\'el temperatures have been obtained with the DIAG method applied to 
double-exchange models\cite{re:dagotto02}.
\indent The Hamiltonian of the system can be written 
as \cite{re:zener52,re:furukawa94}: 
\begin{equation}
{\hat H}=-t\sum_{\langle ij \rangle,\sigma}{\hat c}^\dagger_{i\sigma} {\hat c}_{j\sigma} -
J_H\sum\vec{S}_{i}\cdot\vec{\sigma}_i+J_{AF}\sum_{\langle ij \rangle} \vec{S}_i\cdot \vec{S}_j,
\label{eq:hammanganites}
\end{equation}
\noindent where 
${\hat c}^\dagger_{i\sigma}$ creates a carrier ($e_g$ electron) at site $i$
with spin $\sigma$. The carrier-spin operator interacting
ferromagnetically with the localized Mn-spin $\vec{S}_i$ is
$\vec{\sigma}_i=\sum_{\alpha,\beta}{\hat
c}^\dagger_{i\alpha}\vec{\sigma}_{\alpha,\beta}{\hat c}_{i\beta}$. 
For the manganites, the Hund coupling $J_H$ is large compared to the hopping energy scale $t$ and, as a consequence, suppresses double
occupancy. In this respect, a large $J_H$ coupling acts as a large Hubbard repulsion.
The last term is the  superexchange between localized $t_{2g}$ spins.

Static properties of this model, \eg\ the local spin magnetization and energy of the system have been previously studied with the TPEM on large
lattices.\cite{re:alvarez05} Here the focus will be on more complex observables. Before showing results for dynamical observables we will present first results
for a two-particle static observable, namely, the kinetic energy, $K$. 
A simulation was performed in the ferromagnetic region of Hamiltonian Eq.~(\ref{eq:hammanganites}) with $J_H=4t$ and quarter 
filling. 
The kinetic energy is shown  in Fig.~\ref{fig:kinetic}a for 
a fixed configuration of classical fields (spins, $S_i$, in this case) calculated both with the diagonalization method and with the TPEM for various
values of the cutoff, $M$. The configuration of classical spins was selected after evolving the system with Monte Carlo at $T=0.02t$. 
$K$ converges to the exact value within an error window of less than 2\% for $M\ge100$ which implies that 
the required cutoff is
larger than for one-particle or classical observables.\cite{re:alvarez05}

Periodic boundary conditions will be used throughout the following calculations. The unit of energy is the hopping parameter, $t$.

\subsection{Density of States and Spectral Function}
\begin{figure}[h]
\includegraphics[width=8cm,clip]{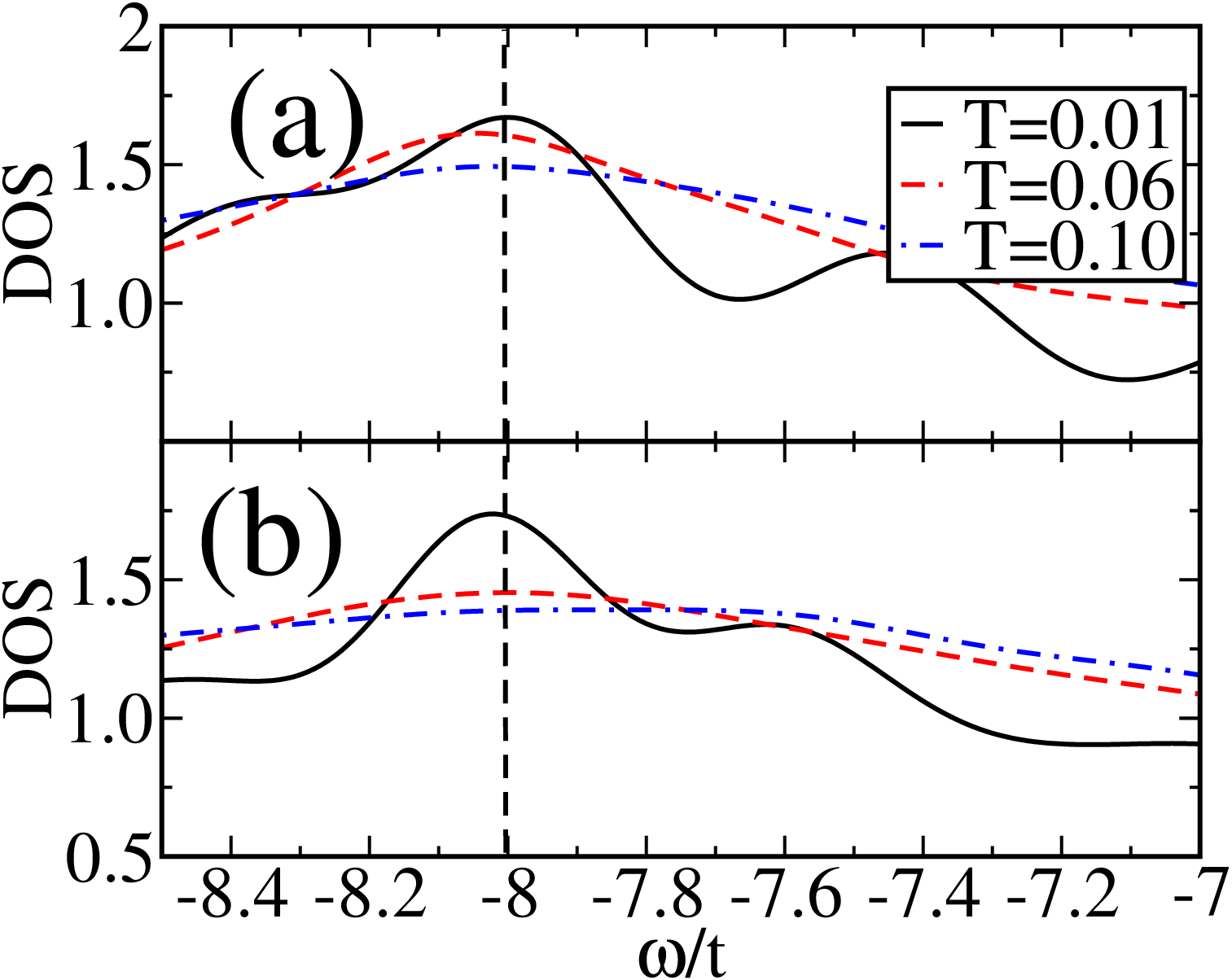}
\caption{DOS at three temperatures
 on a 20$\times$20 lattice calculated with the TPEM and the Monte Carlo
 algorithm, $M=200$, for model Eq.~(\ref{eq:hammanganites}) at $J_H=8t$, $\langle
n\rangle=0.5$ for (a) $J_{AF}=0$ and (b) $J_{AF}=0.02$. The location of the chemical potential is indicated by the vertical dashed line.
\label{fig:dos20x20}}
\end{figure}

\begin{figure}[h]
\includegraphics[width=8cm,clip]{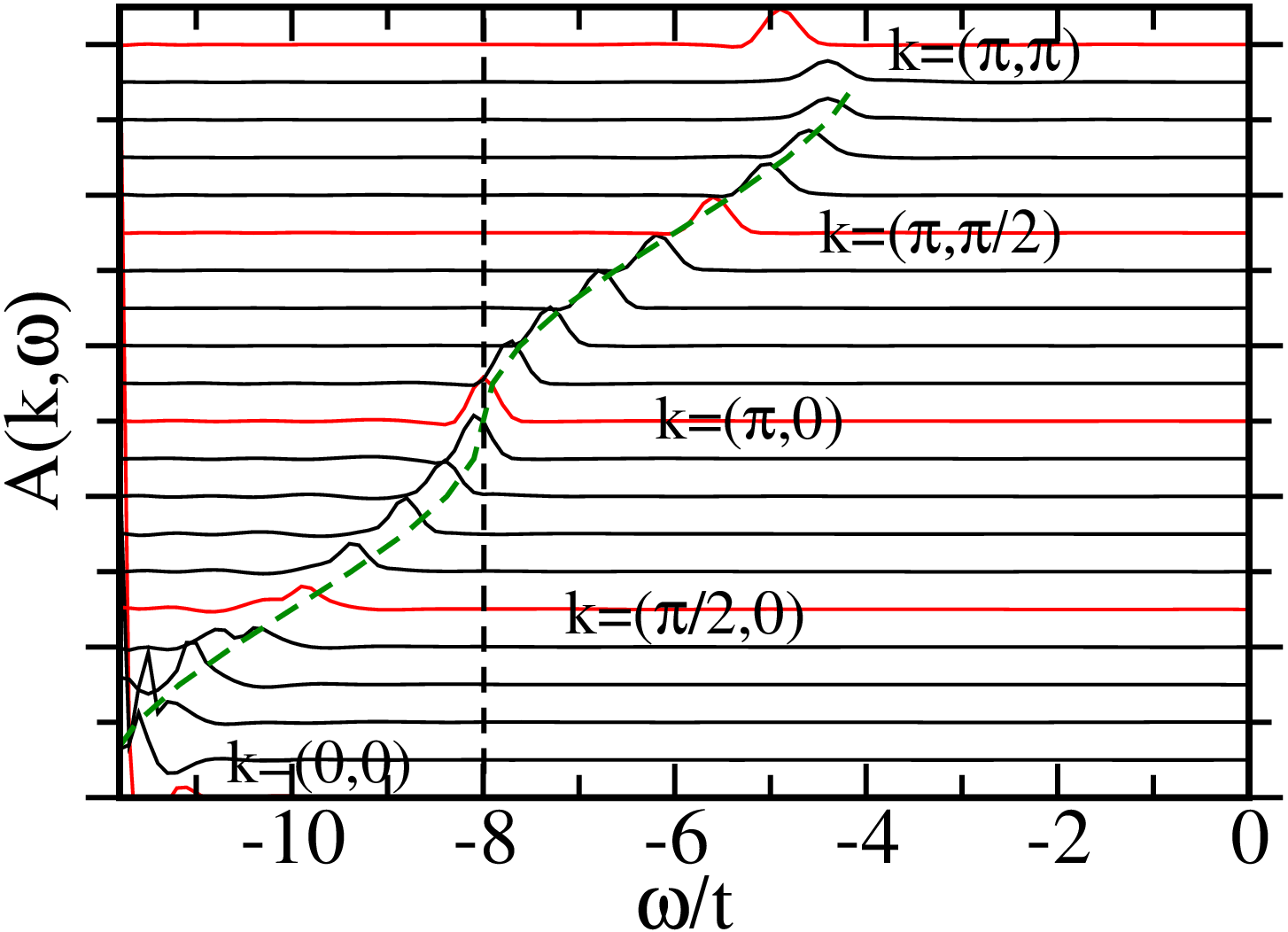}
\caption{Spectral function for the double exchange model calculated on a 20$\times$20 lattice 
with the TPEM at $J_H=8t$, $n=0.5$, $T=0.02t$ (FM system).
The curves are parameterized in terms of $(k_x,k_y)$ from (0,0) to ($\pi$,0) to ($\pi$,$\pi$). 
The location of the Fermi energy is indicated by the vertical dashed line. There is a finite DOS at
the Fermi energy. The dashed line is the spectrum calculated analytically for a perfect ferromagnet.
\label{fig:akw}}
\end{figure}

\begin{figure}[h]
\includegraphics[width=8cm,clip]{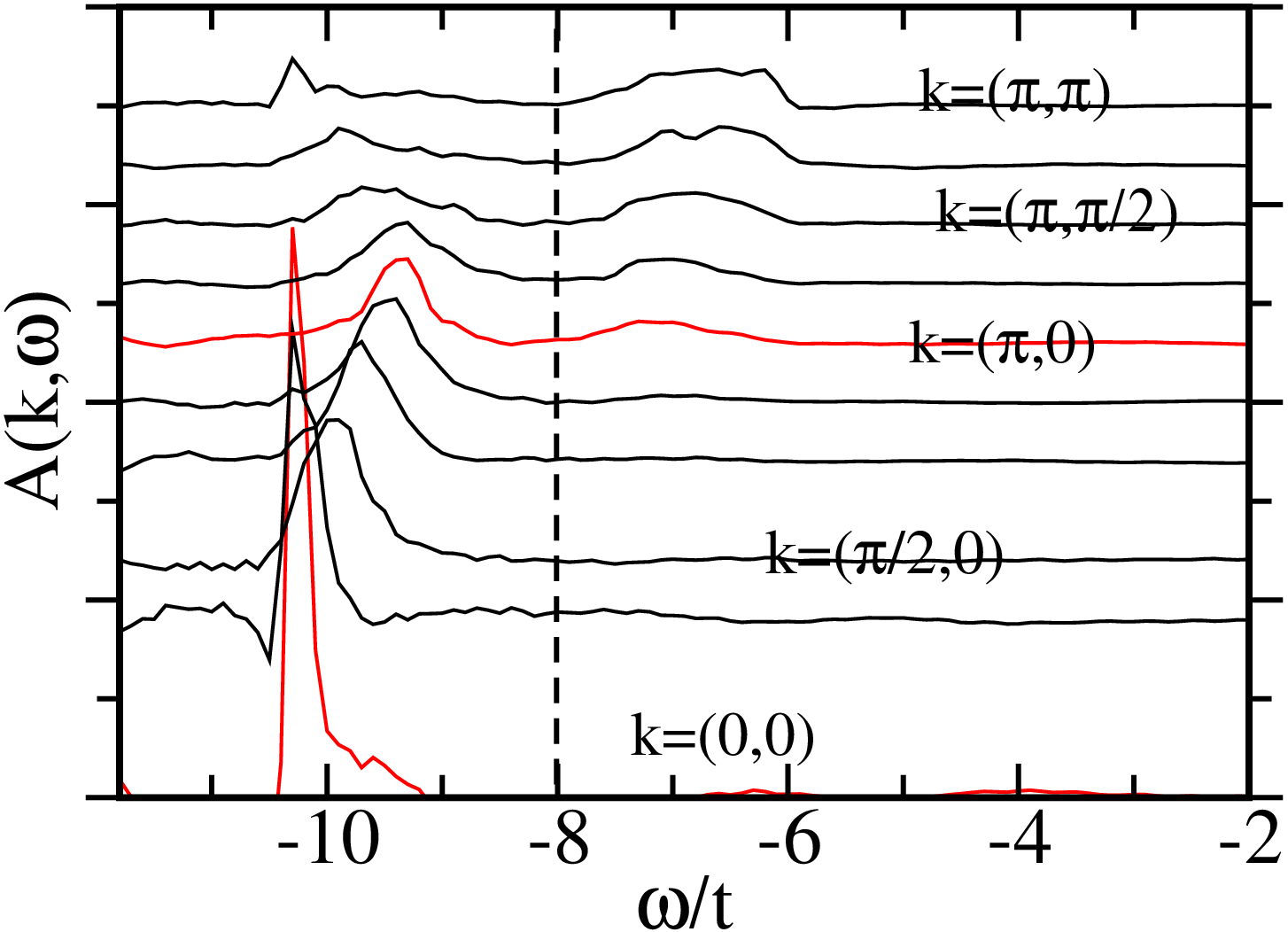}
\caption{For $J_{AF}=0.15$ and the same parameters as in Fig.~\ref{fig:akw}, corresponding
to an insulator, $A(k,\omega)$ shows a clear gap and a more complex form.
The vertical dashed line indicates the position of the Fermi energy.
\label{fig:akw01}}
\end{figure}

We will analyze Eq.~(\ref{eq:hammanganites}) at quarter filling ($\langle n\rangle=0.5$) and $J_H=8t$. For $J_{AF}=0$ the ground state of the system is
ferromagnetic. As $J_{AF}$ is increased the superexchange coupling competes with the double-exchange term and there exists a critical
$J_{AF}$ beyond which the system's ground state is no longer ferromagnetic. This 
critical $J_{AF}$ was found to be $0.05\pm0.01$ in two dimensions and 
$0.12\pm0.01$ in one dimension. Above this value, the structure factor presents peaks at incommensurate momenta 
(and if $J_{AF}$ is large enough the system becomes antiferromagnetic). In the next section we will show that the ferromagnetic region is metallic,
whereas the region with incommensurate momenta is insulating.

First we show that the DOS calculated with the TPEM gives reasonable results by
comparing with the ones obtained with the DIAG (exact) method.
This is presented in Fig.~\ref{fig:kinetic}b for the same parameters as in Fig.~\ref{fig:kinetic}a. The finite
size effects of the DOS are only reproduced by the TPEM with 100 moments or more. Unlike energy integrated quantities, energy dependent
quantities need more moments for convergence, as expected. The DOS in the ferromagnetic phase has two bands centered at approximately
 $\pm J_H$ each with bandwidth equal to $4t$.

Now that we know that the TPEM yields reliable results for the DOS we proceed to
perform Monte Carlo simulations. In Fig.~\ref{fig:dos20x20} the DOS is shown for different temperatures on a 20$\times$20 lattice at $\langle n \rangle=0.5$ and $J_H=8t$. 
The dependence with temperature
is very mild at $J_{AF}=0$ (Fig.~\ref{fig:dos20x20}a) and there is a slight decrease of the DOS at the Fermi energy 
for $J_{AF}=0.02$ (Fig.~\ref{fig:dos20x20}b) which is still 
ferromagnetic. There is no indication of a pseudogap forming at higher temperatures in the ferromagnetic region. However, in photoemission studies for
manganites\cite{re:dessau98} the DOS of metallic samples 
was found to depend more strongly on temperature and to display a more complicated structure than the
one shown in Fig.~\ref{fig:dos20x20}. It is very likely that the inclusion of phonons or many orbitals 
are necessary to be able to
reproduce these features. The influence of these corrections will be prominent mainly near the
critical $J_{AF}$ that separates the metallic from the insulating regions.
This suggests that the relevant regime of our model for typical
 ferromagnetic samples lies very near the metal-insulator transition that occurs with increasing $J_{AF}$.  We will see evidence of this
also when we examine the optical conductivity and conductances in the next section. 

Our tests show that the calculation of the spectral function with Eq.~(\ref{eq:arw}) requires at least $M=100$ for this model.
Using a large value for $M$, $A(k,\omega)$ is plotted in Fig.~\ref{fig:akw} and \ref{fig:akw01} on a $20\times20$ lattice for $J_H=8t$ at 
low temperature
($T=0.02t$) and quarter filling. 
A total of 1000 thermalization steps and 1000 measurement steps were done first with $M=40$. 
An additional 100 steps were used to calculate $A(k,\omega)$ with a large number of moments for the expansion.
For $J_{AF}=0$, Fig.~\ref{fig:akw}, corresponding to a ferromagnet, there is a single peak for every value
of $k$,  as expected since the system is ferromagnetic. 
There is also a clear Fermi surface satisfying $\cos(k_x)+\cos(k_y)=0$ and the peaks follow the dispersion calculated analytically for a
perfect FM.
But for $J_{AF}=0.15$ in Fig.~\ref{fig:akw01}, $A(k,\omega)$
presents a complex form and also a gap opens at the Fermi energy. It is worth noting that the structure factor at this value of $J_{AF}$
 has peaks
at incommensurate momenta, \eg\ $(k_x,k_y)=(\pi,0)$ that are precursors of the antiferromagnetic phase that forms at even 
larger $J_{AF}$.

In all cases, the moments where multiplied by damping factors to 
reduce the high frequency oscillations as explained before.

\subsection{Optical Conductivity}
\begin{figure}[h]
\includegraphics[width=8cm,clip]{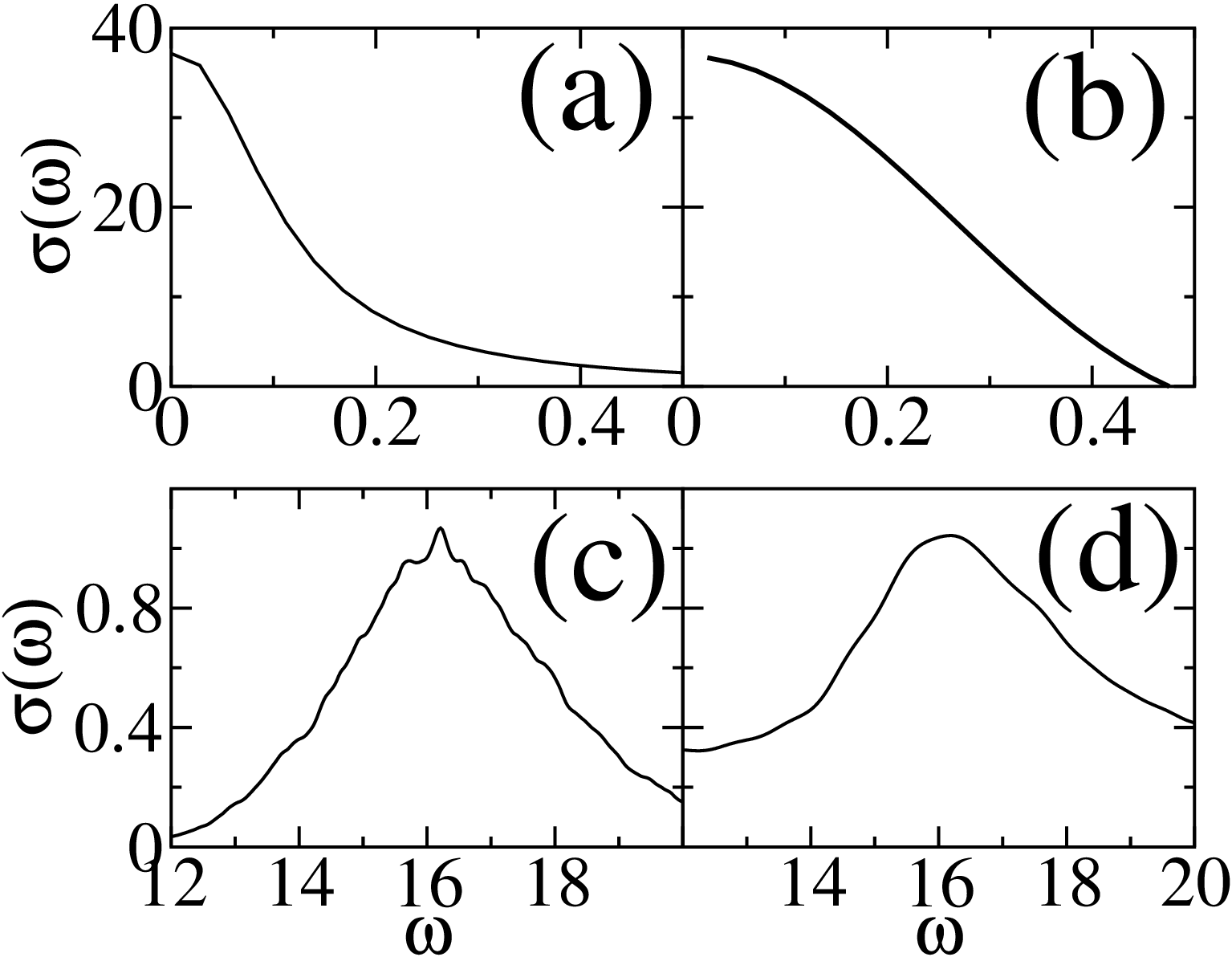}
\caption{Optical conductivity on a $6^3$ lattice calculated with the TPEM and DIAG for the same configurations of classical fields
at $J_H=8t$ and quarter filling. 
(a) $T=0.02t$ calculated
with the DIAG (b) $T=0.02t$ calculated with the TPEM, (c) $T=0.2t$ calculated with
DIAG and (d) $T=0.2t$ calculated with the TPEM. In (b) and (d) $M=200$. \label{fig:optical}}
\end{figure}

\begin{figure}[h]
\includegraphics[width=8cm,clip]{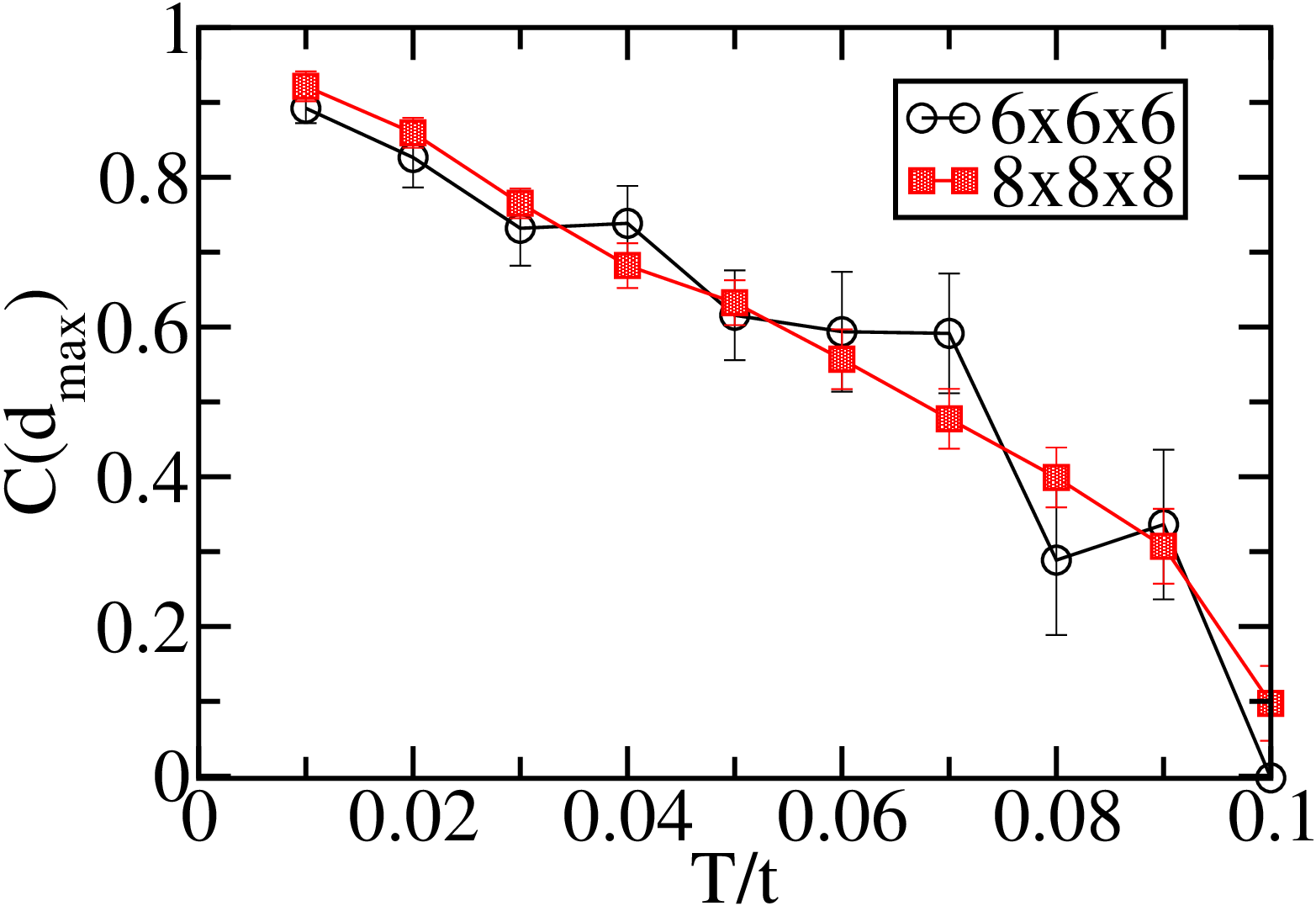}
\caption{Maximum distance spin-spin correlation, $C(d_{max})$, vs.~temperature for the ferromagnetic regime ($J_{AF}=0$)
on $6^3$ and $8^3$ lattices. $d_{max}=\sqrt{3}L/2$ where $L=6$ or $L=8$.\label{fig:correlations}}
\end{figure}

The optical conductivity in units of $e^2/\hbar$ is shown in Fig.~\ref{fig:optical} on a $6^3$ lattice for two configurations 
of classical spins obtained at two different temperatures. 
The figure shows that the TPEM with 200 moments gives results consistent with the diagonalization method.
The value of $T_C=0.09\pm0.01$ was determined from the maximum distance correlation between classical spins 
(Fig.~\ref{fig:correlations}) taken as $distance=L\sqrt{3}/2$, with $L=6$
being the linear length of the lattice. 
Below $T_C$ there is a single peak at $\omega=0$
and above $T_C$ there is an additional peak at $\omega=2J_H$ that increases in intensity as temperature is increased. This peak at finite
$\omega\approx 2J_H$ is due to transitions from both the spins up and down since they have similar occupations above $T_C$. On
the other hand, below $T_C$ the electronic states with spin antiparallel to the $t_{2g}$ spins have high energies and,
therefore,  the transition across the $2J_H$ gap weakens and only the ``Drude'' peak at $\omega=0$ is left.

Fig.~\ref{fig:optical} is in qualitative agreement with the optical conductivity spectra of
 La$_{1-x}$Sr$_{x}$MnO$_3$ (see \eg\ Ref.~\onlinecite{re:okimoto95}) where
the intensity is transfered from low to high energy as temperature increases. Then, our three dimensional 
results suggest that
the spin-fermion model considered here contains the basic physics to reproduce the behavior of the optical
conductivity with temperature at least in the metallic regime. A more accurate treatment would include 
many orbitals, in particular if orbital ordered phases, \eg\ the CE
phase\cite{re:aliaga03}, are relevant\cite{re:kim98,re:kim02}.

\begin{figure}[h]
\includegraphics[width=8cm,clip]{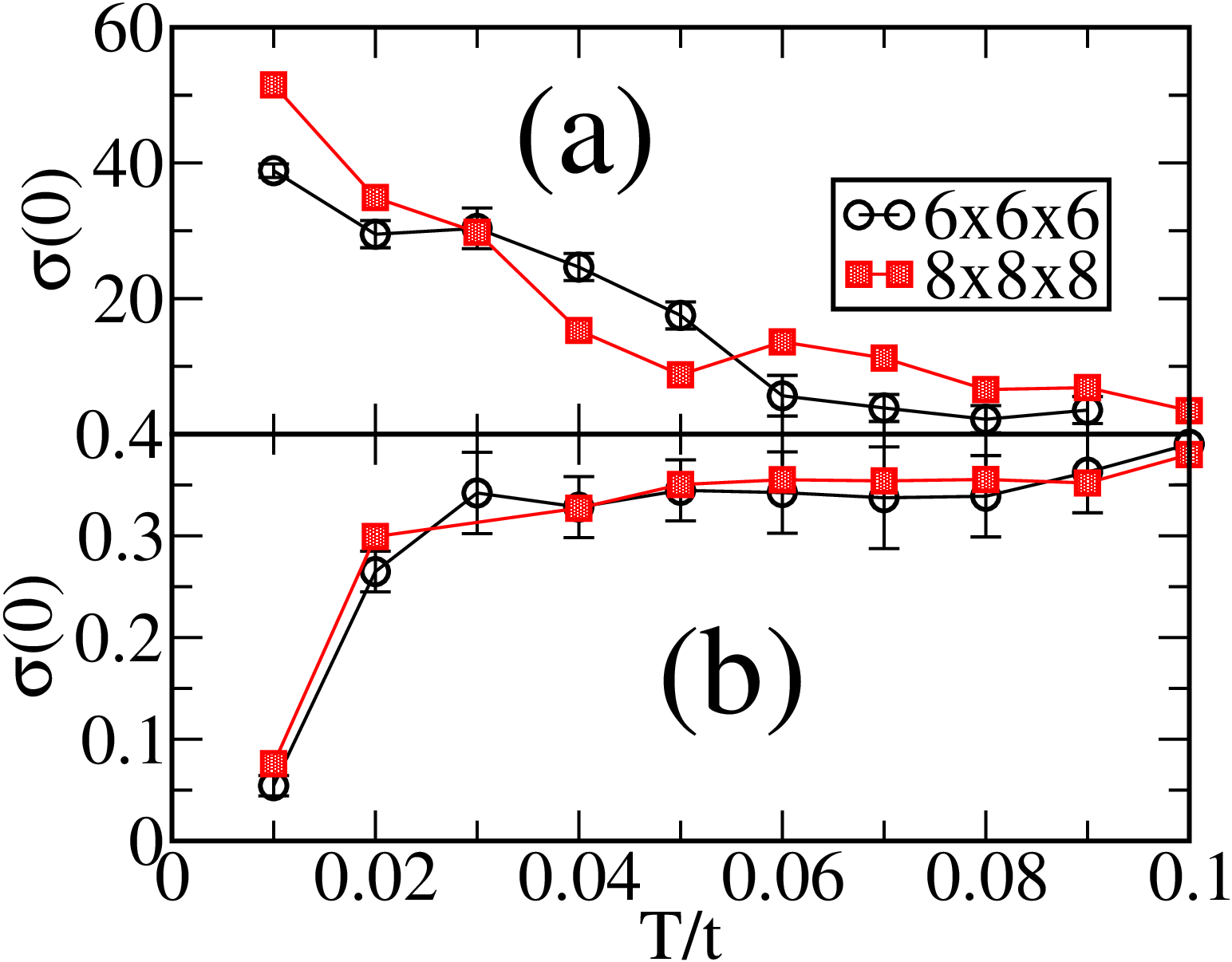}
\caption{$\sigma(0)$ vs.~temperature at $J_H=8t$ and quarter filling on $6^3$ and $8^3$ lattices for 
(a) $J_{AF}=0$ and (b)  $J_{AF}=0.1t$ showing metallic and insulating behavior
with temperature respectively. The results were obtained with the TPEM and $M=200$.
\label{fig:optical8d3jaf0}}
\end{figure}

Now the TPEM  will be used to determine the metal to insulator
transition that occurs in the model when a $J_{AF}\neq0$ is considered in Eq.~(\ref{eq:hammanganites}). This transition was found in
one dimensional finite chains at zero temperature\cite{re:yunoki98}. The quantity of interest is the limit of $\sigma(\omega)$ as 
$\omega\rightarrow0$, 
yielding the conductance $C$. A total of 1000 Monte Carlo steps for thermalization and 1000 measurement steps were performed with $M=40$. 
An additional 100 steps were
used to calculate dynamical properties with a large number of moments for the expansion (usually $M=200$). 
We will examine the three-dimensional case for two values of the superexchange coupling $J_{AF}$.
In Fig.~\ref{fig:optical8d3jaf0}a for $J_{AF}=0$ (corresponding to a ferromagnetic ground state), 
$C$ is plotted as a function of temperature on $6^3$ and $8^3$ lattices with $J_H=8t$ and
$\mu= -8t$, \ie\  for
 a density approximately equal to quarter filling. As expected $C$
 decreases monotonically with increasing temperature.
 On the other hand, for $J_{AF}=0.1t$ the ground state is not ferromagnetic anymore but it is 
 given by states with incommensurate momenta instead. 
 These states are insulating as can be seen from Fig.~\ref{fig:optical8d3jaf0}b. Therefore, the present
 method allows for the analysis of the   metal to insulator transition with increasing $J_{AF}$ in the three dimensional model.
 It was essential to be able to run simulations on $6^3$ and $8^3$ lattices since $4^3$ is known to show strong finite size effects
(\eg\  the low
temperature magnetization increases with temperature in that case). 
Simulations on $8^3$ or more lattice sites are not possible with the DIAG method.

Note that the standard dependence of the resistivity on temperature for
 CMR manganites
 cannot be reproduced with the model given by 
Eq.~(\ref{eq:hammanganites}): the calculated resistivity has no peak at finite temperature, only a monotonous behavior. 
However, there is indication\cite{re:sen04} that 
chemical disorder could transform an insulator into a poor metal at low temperatures. For this to happen the parameters of the 
system ($J_{AF}$
in our model) have to be located very near the border of the metal-insulator separation. 
This is in agreement with the experimental observation\cite{re:tokura96} that CMR manganites in the ferromagnetic region 
have poor conductance and with the results presented
before for the density-of-states and spectral function and provide further evidence that this is the relevant regime to
investigate for possible explanations of the CMR effect. In other words, if small corrections, \eg\ disorder, many orbitals or
weakly-coupled phonons, 
are included far from the critical $J_{AF}$, they will not affect the qualitative trends shown here but they are expected to be 
very important near 
this 
critical $J_{AF}$.

The Hamiltonian parameter corresponding to the superexchange interaction between $t_{2g}$ spins, $J_{AF}$, drives the three-dimensional 
system from a metal to an insulator. In the phenomenological model of Eq.~(\ref{eq:hammanganites}), $J_{AF}$  is an effective
interaction that competes with the double-exchange mechanism. This competition is related to the
tolerance factor of the samples, or the deviation of the Mn-O-Mn angle. Therefore, calculations at 
different values of $J_{AF}$ can simulate the transport and magnetic properties of different 
compounds. For instance, in the Pr$_{1-x}$(Ca$_{1-y}$Sr$_y$)$_{x}$MnO$_3$ system\cite{re:tomioka02} distortions in the
MnO$_6$ octahedra weaken the double-exchange mechanism and the superexchange interaction becomes important as well as collective Jahn-Teller
distortions. Even though in this work we have included only the superexchange coupling, 
the qualitative consequences of this competition can be seen in Fig.~\ref{fig:optical8d3jaf0}.
Moreover, with the formalism presented in this work for the
 TPEM it will be possible
  to obtain linear responses of Hamiltonians that include Jahn-Teller phonons and many orbitals which is not the case with the
 DIAG method because the Hilbert space of these systems is at least twice as 
large as the one for the one-orbital model, \ie\ the computations would be 16 times slower. The complexity of the 
TPEM scales linearly with the
size of the one-electron sector.

\section{Conclusions}\label{sec:conclusions}

The TPEM was extended to include the calculation of two- and four-particle observables without introducing systematic errors 
or uncontrollable approximations.
 By comparing with the exact algorithm 
 it was found that energy-dependent observables need more moments than energy integrated observables but
the results converge with $M\approx 100$ moments. The formulas were applied to a spin-fermion model for manganites on a 20$\times$20
 lattice to analyze 
 the spectral function for the ferromagnetic and incommensurate phases. We concluded that the simple model of Eq.~(\ref{eq:hammanganites})
 cannot reproduce the DOS of typical CMR samples. The inclusion of chemical disorden or weakly-coupled phonons
 would be necessary and these corrections should be important near 
  the border that separates the ferromagnetic from the  region with 
  incommensurate momenta in the phase diagram. 
  Similar conclusions were found by studying  the metal to insulator
transition with increasing superexchange coupling that occurs in the three dimensional model.

The truncated polynomial expansion 
allowed us to calculate dynamical properties of a spin-fermion model for manganites on up to $8^3$ lattices  without approximations. 
Even larger lattices will be reachable with the new generation of supercomputers that can accommodate easily thousands of nodes, as explained before in
\S\ref{subsec:parallel}. Big lattices are particularly necessary in simulations including chemical disorder. Moreover, systems with realistic band
structures\cite{re:schulthess01} will require an increase in the size of the Hilbert space and the calculation of dynamical observables in these 
more realistic models will benefit from the present methods.

\begin{acknowledgments}
We would like to thank Elbio Dagotto for many suggestions to improve the presentation of
this manuscript.
Conversations with H. Aliaga, N. Furukawa, A Moreo, Y. Motome and I. Sergienko 
are also acknowledged.
This research used the SPF computer program and software toolkit developed at ORNL (http://mri-fre.ornl.gov/spf).
Research performed at the Oak Ridge National Laboratory, managed by UT-Battelle, LLC, for the U.S. Department of Energy 
under Contract DE-AC05-00OR22725.

\end{acknowledgments}
\bibliography{thesis}

\end{document}